\documentclass[11pt]{article}
\usepackage{graphicx}
\usepackage{epsfig}
\usepackage{array}
\usepackage[figuresright]{rotating}
\usepackage{amsmath}
\usepackage{amssymb}
\usepackage{amsfonts}

\newcommand{\pibf}{\mbox{\boldmath $\pi$}}
\newcommand{\kappabf}{\mbox{\boldmath $\kappa$}}
\newcommand{\taubf}{\mbox{\boldmath $\tau$}}

\newcommand{\vectau}{\mbox{\boldmath$\tau$}}

\newcommand{\fpartial}{\mbox{$/\!\!\!\partial$}}

\usepackage{cite}
\begin{document}

\begin{center}
{\LARGE{\bf Structure of scalar mesons
 and  the Higgs sector of strong interaction} 
  }\\[1ex] 
Martin Schumacher\\mschuma3@gwdg.de\\
Zweites Physikalisches Institut der Universit\"at G\"ottingen,
Friedrich-Hund-Platz 1\\ D-37077 G\"ottingen, Germany
\end{center}

\begin{abstract}
The scalar mesons $\sigma(600)$, $\kappa(800)$, $f_0(980)$ and $a_0(980)$
together with the pseudo Goldstone bosons $\pi$, $K$ and $\eta$  may be
considered as the Higgs sector of strong interaction. After a long time 
of uncertainty about the internal structure of the scalar mesons there now
seems to be consistency which is in line with the  major parts of
experimental observations. Great progress has been made by introducing the
unified  model of Close and T\"ornqvist. This model states that 
scalar mesons below 1 GeV  may be understood as 
$q^2\bar{q}^2$
in $S$-wave with some $q\bar{q}$ in $P$-wave in the center, further out
they rearrange as $(q\bar{q})^2$ and finally as meson-meson
states. The $P$-wave component inherent in the structure of the 
neutral scalar mesons
can be understood as a doorway state for the formation
of the scalar meson via two-photon fusion, whereas in  nucleon Compton 
scattering these P-wave components serve as intermediate states. The masses
of the scalar mesons are predicted in terms of spontaneous and explicit
symmetry breaking.
\end{abstract}

\section{Introduction}

After more than 30 years of research 
the structures and masses of the 
 $f_0(980)$ and $a_0(980)$ mesons are still a matter of discussion. As a good
candidate for the structure of these mesons the four-quark configuration
$q^2\bar{q}^2$ or $(q\bar{q})^2$ has been introduced   \cite{jaffe76}. 
An other option is the $K\bar{K}$
molecule which has the advantage that twice the $K$-meson mass is approximately
equal to the masses  of the $f_0(980)$ and $a_0(980)$ mesons.  
In case of the
 four-quark configuration  use has been made of the assumption
that diquarks like   $(qq)$ and $(\bar{q} \bar{q})$ experience a binding
between the partners of the diquarks and thus shift the energy of the
meson to the experimentally determined value (for an extensive  list of
references
see e.g. \cite{PDG,schumacher07b,tuan01}). 
Further aspects of the interplay between the $q\bar{q}$ and $q^2\bar{q}^2$
  strutures of scalar mesons may be found in \cite{black00,black01,fariborz09}
  and references therein.

A  unified model of the scalar mesons
above and below 1 GeV  has been introduced by Close and T\"ornqvist
 \cite{close02}. In  this model  the scalar mesons above 1 GeV
are described 
as a conventional $q\bar{q}$ nonet mixed with the glueball of lattice QCD.
Below 1 GeV the states also form a nonet, as implied by the attractive forces
of QCD, but of more complicated nature. Near the center they are 
$(qq)_{\bar{3}}(\bar{q}\bar{q})_3$ in $S$-wave, with some $q\bar{q}$
in  $P$-wave, but  further out they rearrange as  $(q\bar{q})_1(q\bar{q})_1$
and finally as meson-meson states. In the present paper we adopt this model
as a basis but make one essential refinement concerning
the small  $q\bar{q}$ component in  $P$-wave. We follow  Close and T\"ornqvist
 \cite{close02} in considering this component as a minor part of the 
meson structure but give it a new and important interpretation. As a
$q\bar{q}$ structure this $P$-wave component has a large coupling strength
to the two photons in a two-photon fusion reaction producing the 
scalar mesons $\sigma(600)$, $f_0(980)$ and $a_0(980)$. The reason is that 
two photons are capable of producing a $q\bar{q}$ structure component
via the coupling to a quark  loop, whereas the one-step production
of a complicated structure is comparatively weak. This situation resembles
similar cases known in the atomic nucleus, where small components with
a large coupling to the entrance channel are named doorway states. One 
illustrative example is provided by the photoexcitation of the giant-dipole
resonance of the nucleus. In a shell model this excitation corresponds to 
single-particle electric-dipole transitions, whereas the  main structure
of a giant-dipole resonance is a collective motion.

Compton scattering by the nucleon has provided a new access to scalar mesons
below 1 GeV. Since the mesons $\sigma(600)$, $f_0(980)$ and $a_0(980)$ couple
to two photons with parallel planes of linear polarization, they 
provide a $t$-channel contribution to Compton scattering and to the electric
($\alpha$) and magnetic ($\beta$) polarizability of the nucleon. In  recent
works the role of the $\sigma(600)$, the $f_0(980)$ and the $a_0(980)$ 
mesons in nucleon Compton scattering has been
investigated 
\cite{schumacher05,schumacher06,schumacher07a,schumacher07b,schumacher08,schumacher09,schumacher10,schumacher11}.
It has been shown that the $t$-channel
amplitude $A_1(t)$ of nucleon Compton scattering and the $t$-channel
contributions to the polarizabilities $\alpha$ and $\beta$ can be
quantitatively predicted by considering the $\sigma$-meson as a $q\bar{q}$
state. Furthermore, the mass of the $\sigma$ meson was predicted on an 
absolute scale via spontaneous  and explicit symmetry breaking in agreement 
with experimental
information obtained from Compton scattering and the related
polarizabilities. One of these  investigations clearly  showed
that the $\sigma$ meson as a $q\bar{q}$ state has the properties of the 
Higgs boson of strong interaction
\cite{schumacher10}. However, these investigations did not give an
answer to the question  how the $q\bar{q}$ structure of the $\sigma$ meson is 
related to other structures which are suggested by the fact that the 
$\sigma$ meson decays into two pions within a very short lifetime.
Furthermore, the properties and the role of the mesons $f_0(980)$ and $a_0(980)$
have  not been investigated in detail. The purpose of the present study
is to supplement on the previous work by studying properties of
scalar mesons as the Higgs sector of strong interaction.

\section{The $q\bar{q}$, $(q\bar{q})^2$ and dimeson 
structures of scalar mesons}

According to the unified model of Close and T\"ornqvist \cite{close02}  the
$q\bar{q}$, $(q\bar{q})^2$ and dimeson 
structures of scalar mesons are 
simultaneously  part of the structure of the scalar mesons. 
The $q\bar{q}$ structure component is related with the 
entrance channel and the dimeson component with the exit channel. 
The four-quark structures in the form $(qq)(\bar{q}\bar{q})$ or
$(q\bar{q})(q\bar{q})$ serve as the main or central component. 
We will see that it is
not of importance to distinguish  between the two latter four-quark
structures,  because a special force between  diquarks is not essential
for an explanation of the masses of scalar mesons. Rather, it will be shown
that these masses can be understood in terms of spontaneous and explicit
symmetry breaking, where spontaneous symmetry breaking leads to one 
common component of the masses of all scalar mesons independent of the 
flavor structure, whereas the  
differences in mass may  be understood in terms of  explicit
symmetry breaking. This insight brings us closer to the supposition that the
scalar mesons in connection with the pseudo Goldstone bosons may be regarded
as the Higgs sector of strong interaction.

Our first interest is directed to the electrically neutral scalar mesons
$\sigma(600)$, $f_0(980)$ and $a_0(980)$   which on the one hand can be
produced in  two-photon fusions reactions and on the other hand show up
as intermediate states in Compton scattering experiments by the nucleon. 
For the discussion  we  start with an  ansatz for the structures of the
three non-charged scalar mesons in the following form
\begin{eqnarray}
&&\sigma=\frac{u\bar{u}+d\bar{d}}{\sqrt{2}}\leftrightarrow u\bar{u}d\bar{d}
\leftrightarrow \pi\pi, \label{struc1}\\
&&f_0\approx\frac{1}{\sqrt{2}}\left( \frac{u\bar{u}+d\bar{d}}{\sqrt{2}}-
s\bar{s}\right) \leftrightarrow  \frac{s\bar{s}(u\bar{u}+d\bar{d})}{\sqrt{2}}
\leftrightarrow \pi\pi,K\bar{K}, \label{struc2}\\
&&a_0\approx\frac{1}{\sqrt{2}}\left( \frac{-u\bar{u}+d\bar{d}}{\sqrt{2}}+
s\bar{s}\right) \leftrightarrow s\bar{s}  \frac{(u\bar{u}-d\bar{d})}{\sqrt{2}}
\leftrightarrow \eta\pi,K\bar{K} \label{struc3}
\end{eqnarray}
which of course needs a detailed justification. The first configuration
 is the $q\bar{q}$ P-wave part of the central ``core'' 
state of the scalar meson, the second one possible version of a
$(q\bar{q})(q\bar{q})$ state which easily can be rearranged into a
$(qq)(\bar{q}\bar{q})$ configuration. For sake of convenience we will
write  $(q\bar{q})^2$  in the following for both configurations.
These  $(q\bar{q})^2$ configurations may be
considered as the main components  of the central ``core'' state.
The third configurations represent the dimeson states observed in the exit
channels. First we notice 
that the transition from the $q\bar{q}$ configurations to the
$(q\bar{q})^2$ configurations
is possible by a rearrangement of the quark structure without a 
$q\bar{q}$ pair creation of annihilation. This makes them a natural
$^3P_0$ partner of the main $(q\bar{q})^2$ structures. Later on we
will see that these $q\bar{q}$ structures are compatible with the transition
amplitudes ${\cal M}(M\to \gamma\gamma)$  observed for the three scalar mesons
$M$ in two-photon fusion reactions as well as in Compton scattering
experiments.
In (\ref{struc3}) 
the minus sign attached to the $\bar{u}$ quark in the $q\bar{q}$ configuration
 follows from the sign convention of Close \cite{close79}. 
This minus sign is also present in the flavor wave-function of the $\pi^0$
 meson which has a strong impact on the differential cross section for Compton
 scattering by the nucleon and, therefore, can be investigated with high
 precision.  
As shown in \cite{schumacher07b} this sign convention  has the advantage 
that  the signs of $t$-channel Compton scattering amplitudes are 
predicted in agreement with the experimental observation. Of course,
also the opposite sign in the $q\bar{q}$ structures of the $\pi^0$ and 
$a_0(980)$ 
meson would be possible if we reverse the sign of the isospin operator
$\taubf$ as explained in \cite{schumacher07b} and proposed in \cite{halzen84}.
The $q\bar{q}$ configuration  of the $a_0(980)$ meson violates isospin
conservation. 
This is of no problem because we consider the $q\bar{q}$ configuration 
as a doorway state which is coupled to two photons on the one side  and 
on the other side to the main configuration of the meson  core
via rearrangements of the quark structure. For the coupling to two photons
the mixing of two isospins is allowed because in an electromagnetic transitions
the isospin may change by $\Delta I=0,\,\pm 1$, so that the first term and the 
second term in  the $q\bar{q}$ configuration  of the $a_0(980)$ meson
can be exited simultaneously.
In the charged $a_0^{\pm}(980)$ mesons the isospin violating doorway state is
not of relevance so that only the isospin conserving 
dominant $(q\bar{q})^2$  
 structure components  have  to be taken into account.

Table \ref{qqqq} summarizes the $(q\bar{q})^2$ configurations 
of the scalar nonet (see \cite{jaffe76}).
\begin{table}[h]
\caption{Summary of scalar mesons in the $(q\bar{q})^2$ representation
according to \cite{jaffe76}. $Y$: hypercharge, $I_3$: isospin component, 
$f_s$: fraction of strange
and/or antistrange quarks in the tetraquark structure.}
\begin{center}
\begin{tabular}{l|ccccc|cc}
\hline
$\,\,\,\,Y\,\,\backslash\,\, I_3$& $-1$ & $-1/2$ & $0$ & $+1/2$ & $+1$
&meson&$f_s$\\
\hline
$+1$ & &$d\bar{s}u\bar{u}$& &$u\bar{s}d\bar{d}$&&$\kappa(800)$&1/4\\
\,\,\,\,\,0 &&& $u\bar{d}d\bar{u}$&&&$\sigma(600)$&0\\
\,\,\,\,\,0 & $d\bar{u}s\bar{s}$ & & $s\bar{s}(u\bar{u}-d\bar{d})/\sqrt{2}$&& 
$u\bar{d}s\bar{s}$&$a_0(980)$&1/2\\
\,\,\,\,\,0 &&& $s\bar{s}(u\bar{u}+d\bar{d})/\sqrt{2}$ &&&$f_0(980)$&1/2\\
$-1$  & &$s\bar{u}d\bar{d}$& &$s\bar{d}u\bar{u}$&&$\bar{\kappa}(800)$&1/4\\
\hline
\end{tabular}
\end{center}
\label{qqqq}
\end{table}
As frequently emphasized, the different numbers of strange quarks in the
$(q\bar{q})^2$ structures  may be used for an explanation of the 
different masses  of the scalar mesons. The $\sigma$-meson has no strange
quark and, therefore, the smallest mass. The members of the $\kappa(800)$ 
meson-quartet
have one strange quark in the $(q\bar{q})^2$ configuration 
and, therefore, an
intermediate mass. The members of the ($a_0(980),f_0(980)$)  meson-quartet 
have  two strange
quarks in the $(q\bar{q})^2$ configuration and, therefore, the largest 
mass.

\section{The doorway mechanism applied to scalar mesons}

The doorway mechanism has been developed in nuclear physics to describe 
the excitation of complex structures in nuclei via the excitation of
simple structures. We apply this approach to particles, especially here to
neutral scalar mesons.
The formal ansatz is as follows (see \cite{feshbach67})
\begin{equation}
\psi=\psi_1+\psi_2+\psi_3,
\label{doorwayansatz}
\end{equation}
where the total wave function is written as the sum of three
terms. Furthermore, it is assumed that the Hamiltonian has matrix elements
between $\psi_1$ and $\psi_2$, between $\psi_2$ and $\psi_3$, but not between
$\psi_1$ and $\psi_3$. 
The state   $\psi_2$ is orthogonal to
$\psi_1$, and  $\psi_3$  orthogonal to both  $\psi_1$ and  $\psi_2$.
We interpret the three terms in the following way. 
The state $\psi_1$ is the ground state, $\psi_2$ the doorway state and
$\psi_3$ the main complex structure of the meson. For our case this means
$\psi_1=|0\rangle$, $\psi_2=|q\bar{q}\rangle$ and 
$\psi_3=|(q\bar{q})^2\rangle$. Then the amplitude for the 
$\gamma\gamma\to (q\bar{q})^2$ transition may be written in the generic 
form
\begin{equation}
A_{\gamma\gamma\to (q\bar{q})^2}\propto\frac{\langle 0|H'_1|q\bar{q}\rangle
\langle q\bar{q}|H'_2|(q\bar{q})^2\rangle}{s_0-s},
\label{amp0}
\end{equation}
where in the nominator the first matrix element  corresponds to 
the transition from the ground state into the doorway state and the second 
matrix element to the transition from the doorway state into the main complex
structure of the meson which here is represented in terms of a
$(q\bar{q})^2$ configuration.

Hypothetically, we may assume that the $(q\bar{q})^2$ configuration
is a non-decaying state. In this case the quantity $s_0$ may be identified
with a definite square of a mass  of the particle which commonly is denoted  
as the bare mass $m_0$. The bare mass corresponds to a 
bare propagator, being  of the form
\begin{equation}
P(s)=\frac{1}{m^2_0-s}
\label{prop0}
\end{equation}  
with a pole on the real axis (see e.g. \cite{tornqvist95,boglione02}),
corresponding to a non-decaying state. 

The propagator given in (\ref{prop0}) has the disadvantage that it 
does not take into account the two-photon decay which is possible
even when the decay of the $(q\bar{q})^2$ configuration  into
two  mesons does not take place. In this case the time-dependent state
of the particle may be described in the form
\begin{equation}
\psi^0(t)=\psi^0_0 e^{-i(m_0-\frac12 \,i\,\Gamma_{\gamma\gamma})t}
\label{state}
\end{equation} 
corresponding to the propagator
\begin{equation}
P(s)=\frac{1}{m^2_0-\frac14\Gamma^2_{\gamma\gamma}-i\,m_0\Gamma_{\gamma\gamma}
-s}.
\label{prop0a}
\end{equation}
This latter refinement has to be kept in mind when we make use of the
approximation given in   (\ref{prop0}) in the following. 
The 
pole corresponding to the propagator in (\ref{prop0a}) 
is located in lower half of the  $\sqrt{s}$-plane at
$\sqrt{s_0}=m_0-\frac12 \,i\,\Gamma_{\gamma\gamma}$
or on the second Riemann 
sheet of the $s$-plane \cite{roman65}.

\subsection{Structure and two-photon width of the $\sigma$ meson}

The $\sigma$-meson is a strongly decaying particle so that the decay into two
pions cannot be simply  taken into account by replacing the 
two-photon width $\Gamma_{\gamma\gamma}$ in (\ref{prop0a}) by a constant 
total width $\Gamma_{\rm tot}$. It rather has been proposed to introduce
a vacuum polarization function  \cite{tornqvist95,boglione02}
$\Pi(s)$ which accounts for all the contributions to the propagator
$P(s)$. The imaginary part of $\Pi(s)$ may be obtained from unitarity
considerations. Since the vacuum polarization function, $\Pi(s)$, is an
analytic function, its real part can be deduced from the imaginary part by
making use of a dispersion relation. At this point, we can write the
propagator in terms of the vacuum polarization function:
\begin{equation}
P(s)=\frac{1}{m^2_0 + \Pi(s) -s}=\frac{1}{m^2(s)-s -i\,m_{\rm BW}
\Gamma_{\rm tot}(s)},
\label{prop2}
\end{equation}  
having identified 
\begin{equation}
\Gamma_{\rm tot}(s)=-\frac{{\rm Im}\Pi(s)}{m_{\rm BW}}
\label{prop3}
\end{equation}
and
\begin{equation}
m^2(s)=m^2_0+{\rm Re}\Pi(s).
\label{prop4}
\end{equation}
Here $m^2(s)$ is the  running squared mass, given by the sum of the bare
mass squared  and the real part of the vacuum polarization function 
${\rm Re}\Pi(s)$, which is responsible for the {\it mass} shift.
The imaginary part of the vacuum polarization function ${\rm Im}\Pi(s)$ 
is directly proportional to the width of the state. The mass shift function
${\rm Re}\Pi(s)$ is generally negative and is approximately constant in the
energy regions far from any threshold. The Breit-Wigner mass $m_{\rm BW}$
entering into (\ref{prop2}) and (\ref{prop3}) is defined by the relation
 \cite{tornqvist95,boglione02}
\begin{equation}
m^2(s)-s=m^2_0+{\rm Re}\Pi(s)-s=0
\label{bw}
\end{equation}
i.e. by the  intersection point of the running square mass with the variable
$s$. This leads to the definition of the Breit-Wigner mass
\begin{equation}
m^2_{BW}=m^2_0+{\rm Re}\Pi(m^2_{BW}).
\label{bw1}
\end{equation}

The final goal is to describe the $\sigma$ meson in terms of a particle 
with a particle mass $M_R$ and a mean lifetime $1/\Gamma_R$.
The mass $M_R$ and the width $\Gamma_R$ of this decaying particle are 
determined, in a process
independent way, by the pole of the propagator. The reason for this property
is that a decaying particle can be represented in the form
\begin{equation}
\Psi(t)=\Psi_0 e^{-i\,(M_R-\frac12\, i\, \Gamma_R) t}.
\label{psi}
\end{equation}
Dispersion theory shows (see e.g. \cite{roman65}) that the complex mass in
the exponent of (\ref{psi}) corresponds to a pole in lower half of the
$\sqrt{s}$-plane, or equivalently, to a pole on the second $s$-sheet.
Consequently, in order to
find  the pole position, we have to continue Eq. (\ref{prop2})
into the complex s-plane onto the second  $s$-sheet. 
Then the  propagator may be written in the form
\begin{equation}
P(s)=\frac{1}{s_R-s}
\label{prop4}
\end{equation}
where $\sqrt{s_R}=M_R-i\Gamma_R/2$.
For the $\sigma$ meson the pole is found on the second sheet
at the pole position
$\sqrt{s_\sigma}=M_\sigma-i\,\Gamma_\sigma/2$ with $M_\sigma=441^{+16}_{-8}$ 
MeV and $\Gamma_\sigma=544^{+18}_{-25}$ MeV \cite{caprini06}.

Details of this procedures have been described in a large number of recent
publications and applied to extract the two-photon width
$\Gamma_{\gamma\gamma}$ of the $\sigma$ meson (for a list of references
see \cite{schumacher10}). Therefore, it is not necessary
to give more information here.
Following the notation of Oller et al. \cite{oller08a,oller08b} the two-photon
decay width of the $\sigma$ meson is given by
\begin{equation}
\Gamma(\sigma\to
\gamma\gamma)=
\frac{|g_{\sigma\gamma\gamma}|^2}{16\pi M_\sigma},
\label{width1}
\end{equation}
where $g_{\sigma\gamma\gamma}$ is
the residue at the pole $s_R$. 

Experimentally Eq. (\ref{width1}) is investigated by carrying out  dispersive
theoretical studies of the reactions $\gamma\gamma\to\pi^0\pi^0$,
$\gamma\gamma\to \pi^+\pi^-$, and of  pion scattering data. 
A complete list of recent results obtained 
in this way is given in \cite{schumacher10} leading to an average result for
the  two photon width of
\begin{equation}
\Gamma(\sigma\to\gamma\gamma)=(2.3\pm 0.4)\, {\rm keV}.
\label{twopion1}
\end{equation}
Because of the consistency of the large number of recent evaluations this
result may be considered as a reliable value. 

\subsection{Observation of the non-decaying $\sigma$ meson via Compton
  scattering by the nucleon}

It is of interest to  point out how the bare mass $m_0$
of the non-decaying meson enters into the $t$-channel part of the amplitude
for Compton scattering
and apply this  to the $\sigma$ meson in the first place.
For this purpose Eq. (\ref{amp0}) may be compared 
with the amplitude for $t$-channel Compton scattering being
\begin{equation}
A_{\gamma\gamma}=\frac{\langle 0|H'_1|q\bar{q}\rangle
\langle q\bar{q}|H'_3|N\bar{N}\rangle}{t-m_0}\sin^2\frac{\theta}{2}\equiv
\frac{{\cal M}(M\to \gamma\gamma)g_{MNN}}{t-m_0}\sin^2\frac{\theta}{2},
\label{amp2}
\end{equation} 
where $\theta$ is the c.m. scattering angle of Compton scattering
\cite{schumacher05}. In (\ref{amp2}) the transition 
amplitude $\langle 0|H'_1|q\bar{q}\rangle$ is the same as in 
(\ref{amp0}) and has been identified with the transition
amplitude ${\cal M}(M\to \gamma\gamma)$ of the meson $M$ to two
photons. The transition 
amplitude $\langle q\bar{q}|H'_2|(q\bar{q})^2\rangle$ has been replaced
by the relevant transition amplitude
$\langle q\bar{q}|H'_3|N\bar{N}\rangle$ and the latter has been identified
with the meson-nucleon coupling constant $g_{MNN}$. The kinematical case
of backward Compton scattering $\theta=\pi$ corresponds to the two-photon
fusion reaction with vanishing 3-momentum transfer, i.e. ${\bf k}_1+{\bf
    k}_2=0$. In this case we have $\sin^2\frac{\theta}{2}=1$. For smaller
  scattering angles there are kinematical constraints which are taken into
  account by the factor $\sin^2\frac{\theta}{2}$. 
We see that in (\ref{amp2}) there are no
effects of a vacuum polarization function. The reason is that the
$N\bar{N}$ pair production process takes place in the unphysical region.
Or, in other words, in case of Compton scattering the $\sigma$ meson is in the
status of a non-decaying particle.

As will be discussed in more detail in the next subsection the transition
amplitude of the $\sigma\to\gamma\gamma$ decay may be calculated via
\begin{equation}
{\cal M}(\sigma\to \gamma\gamma)=\frac{\alpha_e}{\pi f_\pi}N_c\left[\left(\frac23
\right)^2+\left(-\frac13\right)^2\right]=\frac53\frac{\alpha_e}{\pi f_\pi}
=41.9\times 10^{-6}\,{\rm MeV}^{-1}
\label{width2}
\end{equation}
where $N_c=3$, $\alpha_e=1/137.04$ being the fine-structure constant and 
$f_\pi=(92.42\pm0.26)$ MeV the pion decay constant. In the quark-level
linear $\sigma$ model (QLL$\sigma$M) \cite{delbourgo95} (see also
\cite{schumacher06,beveren09} and references therein)
the bare mass of the $\sigma$ meson is predicted in the
form 
\begin{equation}
m_\sigma=\left(\frac{16\pi^2}{3}f^2_0 + \hat{m}^2_\pi \right)^\frac12=
666\, {\rm MeV}
\label{width3}
\end{equation}
where the average pion mass is $\hat{m}_\pi= 138$ MeV and the pion decay
constant in the chiral limit $f_0=89.8$ MeV \cite{nagy04}
have been inserted. Using these
numbers we arrive at the two-photon width of the $\sigma$ meson
\begin{equation}
\Gamma(\sigma\to\gamma\gamma)=\frac{m^3_\sigma}{64\pi}|{\cal
  M}(\sigma\to\gamma\gamma)|^2 =2.6 \,{\rm keV}.
\label{width4}
\end{equation}
which has to be compared with 
\begin{equation}
\Gamma(\sigma\to\gamma\gamma)=(2.6\pm 0.3)\, {\rm keV}
\label{width5}
\end{equation}
obtained from the $t$-channel part of the electric polarizability $\alpha_p$
of the proton \cite{schumacher10}. The error given in (\ref{width5})
corresponds to the 10\% precision of the experimental electric polarizability
$\alpha_p$.
The only theoretical result entering into the analysis leading to the value
given in (\ref{width5}) is the QLL$\sigma$M prediction $m_\sigma=666$ MeV
of the bare mass of the $\sigma$ meson.

The result for the  two-photon width of the $\sigma$ meson given in Eqs. 
(\ref{width4}) and (\ref{width5}) is based on the supposition
that the $\sigma$ mesons entering into the Compton scattering amplitude
is identical with the bare, i.e. non-decaying particle. Apparently, this
supposition has been confirmed experimentally with high precision. 
Furthermore, it has been
confirmed that the two-photon excitation of the $\sigma$ meson proceeds
through the $q\bar{q}$ doorway configuration.

The content of this section is  important for the following reasons.
First it has been made transparent how the $\sigma$ meson observed
as a broad resonance in a two-photon fusion reaction is related to the $\sigma$
meson of definite mass $m_0$ showing up as an intermediate state in 
Compton scattering. The experimentally verified identity of the
transition matrix elements ${\cal M}(\sigma\to \gamma\gamma)$ in the two cases,
i.e. Compton scattering and two-photon fusion reaction leading to the bare
particle,
has been traced back to the fact that the doorway state of the two-photon
fusion reaction is identical with the intermediate state of Compton
scattering.

\subsection{Two-photon widths and doorway-structure of scalar mesons}

The  $q\bar{q}$ configurations adopted 
to describe the structure of the doorway states of a scalar meson $M$ are
acceptable only when they are capable of predicting  the experimental
two-photon widths $\Gamma(M\to \gamma\gamma)$ of the meson.

The electromagnetic properties of the $q\bar{q}$ structures 
of the scalar mesons has been investigated 
in previous papers \cite{schumacher07b,schumacher08} so that it is only
necessary here to update the previous arguments. For the discussion 
of the electromagnetic structures it is of major advantage to compare 
pseudoscalar and scalar mesons with each other.
For pseudoscalar mesons  having the constituent quark structure
\begin{equation}
|q\bar{q}\rangle=a|u\bar{u}\rangle+b|d\bar{d}\rangle+c|s\bar{s}\rangle,
\quad a^2+b^2+c^2=1,
\label{Eq-d}
\end{equation}
the two-photon decay amplitude may be given in the form 
\cite{scadron04,schumacher07b}
\begin{equation}
{\cal M}(P\to\gamma\gamma)
=\frac{\alpha_e}{\pi f_\pi}N_c\sqrt{2}\langle\, e^2_q\rangle,\quad
\mbox{with} \quad  \langle  e^2_q \rangle 
= a\, e^2_u +b\, e^2_d +c\,(\hat{m}/m_s)\,e^2_s,
\label{M-formula}
\end{equation}
where $\hat{m}$ is the average constituent mass
of the light quarks and $m_s$ the constituent mass of the strange quark.
Numerically we have $m_s/\hat{m}´\simeq 1.44$ \cite{scadron04,scadron06}. 

In case of scalar mesons the same result is obtained except for the effects
of the kinematical factor (see \cite{schumacher07b} and references therein)
\begin{equation}
\epsilon_{2\mu}\epsilon_{1\nu}(g^{\mu\nu}\,k_2\cdot k_1 - k^\mu_1\,k^\nu_2)
\label{scalarfactor}
\end{equation}
contained in the respective decay amplitude. This kinematical factor
replaces the corresponding factor
\begin{equation}
\epsilon_{\mu\nu\alpha\beta}\,\epsilon^{*\mu}_1\,k^{\nu}_1
\epsilon^{*\alpha}_2\,k^\beta_2
\label{pseudofactor}
\end{equation}
valid for pseudoscalar mesons. Numerically these factors are the same,
except for the fact that they distinguish between the two cases of linear
polarization of the two photons, i.e. perpendicular planes of linear
polarization in case of pseudoscalar mesons and parallel planes of linear
polarization in case of scalar mesons (see \cite{schumacher07b} for details).
This difference in the kinematical factors leads to a
 correction factor 
\begin{equation}
V_q(\xi)=2\xi[2+(1-4\xi)I(\xi)],
\label{twophotonsigma-1}
\end{equation} 
which enters as a multiplicative factor  in (\ref{M-formula}). The quantity 
$\xi$ is given by 
$\xi= m^2_q/m^2_M$, with $m_q$ being  the constituent quark mass and 
$m_M$  the meson mass.
The quantity  $I(\xi)$ is the triangle loop integral given in \cite{beveren09}
\begin{figure}[h]
\centering\includegraphics[width=0.6\linewidth]{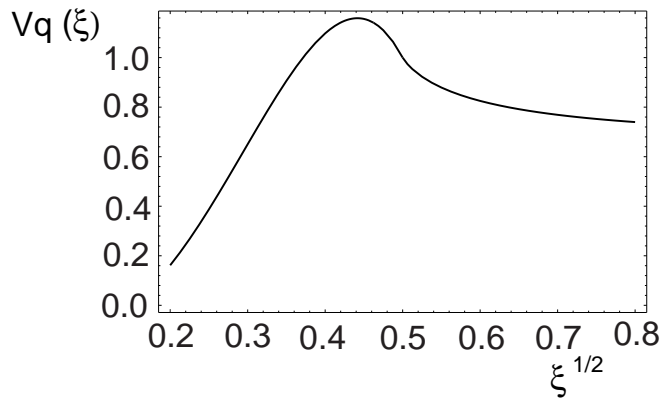}
\caption{Triangle loop integral $V_q(\xi)$ versus $\xi^{1/2}=m_q/m_M$,
where $m_q$ is the constituent quark mass and $m_M$ the meson mass.} 
\label{vq}
\end{figure}
\begin{equation}
I(\xi)
\begin{cases}
=\frac{\pi^2}{2}-2\log^2\left[\sqrt{\frac{1}{4\xi}}+
 \sqrt{\frac{1}{4\xi}-1}\right]+ 2\pi\,i\,\log\left[\sqrt{\frac{1}{4\xi}}+
\sqrt{\frac{1}{4\xi}-1}\right]\,\,\,(\xi\leq 0.25),
\\
=2\arcsin^2\left[\sqrt{\frac{1}{4\xi}}\right]\,\,\,(\xi\geq 0.25).
\end{cases}
\label{twophotonsigma-2}
\end{equation}
The correction factor $V_q(\xi)$ is depicted in Figure \ref{vq}.
From Figure \ref{vq} it can be obtained that the correction factor amounts to
$V_q(\xi)=1$ for $\sqrt{\xi}=0.373$ and  $\sqrt{\xi}=0.5$. For the 
$\sigma$ meson  $\sqrt{\xi}=0.5$ is fulfilled to a very good approximation
so that $V_q(\xi)=1$ may be used. For the $f_0(980)$ and $a_0(980)$ mesons
the constituent quark mass is expected to be between 330 MeV and 490 MeV, i.e.
between the non-strange prediction and one-half of the meson mass,
so that according to Figure \ref{vq}  the possible correction factor $V_q$ 
may be disregarded
in view of other possible   uncertainties.
This consideration justifies that Eq. 
(\ref{M-formula}) may be used for pseudoscalar mesons as well as for
scalar mesons.

Using      
(\ref{M-formula}) for pseudoscalar mesons as well as scalar mesons
and adjusting the two-photon widths
\begin{equation}
\Gamma({M\to\gamma\gamma})=\frac{m^3_M}{64 \pi}|{\cal M}(M\to\gamma\gamma)|^2
\label{twophoton}
\end{equation}
to the experimental data 
we arrive at the 
$|q\bar{q}\rangle$ structures of pseudoscalar and scalar mesons as given in
Table \ref{structure}.
\begin{table}[h]
\caption{$q\bar{q}$-structures of scalar and pseudoscalar mesons calculated
  from the experimental two-photon decay width $\Gamma({M\to\gamma\gamma})$}
\begin{center}
\begin{tabular}{llll}
\vspace{1mm}
meson &  $q\bar{q}$ structure & $\Gamma({M\to\gamma\gamma})$ [keV]
&Reference\\
\hline\vspace{1mm}
$|\pi^0\rangle$&$\!\!\!\!\!=|IV\rangle$&$(7.74\pm 0.55)\times 10^{-3}$
&\cite{schumacher06}\\
\vspace{1mm}
$|\eta\rangle$&$\!\!\!\!\!=\frac{1}{\sqrt{2}}(1.04\,|IS\rangle-0.96\,|s\bar{s}
\rangle)$&$0.510\pm 0.026$&\cite{PDG}\\\vspace{1mm}
$|\eta'\rangle$&$\!\!\!\!\!=\frac{1}{\sqrt{2}}(0.83\,|IS\rangle+1.15\,|s\bar{s}
\rangle)$&$4.29\pm0.15$&\cite{PDG}\\\vspace{1mm}
$|\sigma(666)\rangle$&$\!\!\!\!\!=|IS\rangle$&$2.6\pm 0.3$&
\cite{schumacher10}\\ \vspace{1mm}
$|f_0(980)\rangle$&$\!\!\!\!\!=\frac{1}{\sqrt{2}}(0.52\,|IS\rangle-1.31\,
|s\bar{s}
\rangle)$&$0.29^{+0.07}_{-0.08}$&\cite{PDG}\\\vspace{1mm}
$|a_0(980)\rangle$&$\!\!\!\!\!=\frac{1}{\sqrt{2}}(0.83\,|IV\rangle+1.15
\,|s\bar{s}
\rangle)$&$0.30\pm 0.10$&\cite{PDG}\\ \vspace{1mm}
$|IS\rangle$&$\!\!\!\!\!=\frac{1}{\sqrt{2}}(|u\bar{u}\rangle+|d\bar{d}
\rangle$&
\\\vspace{1mm}
$|IV\rangle$&$\!\!\!\!\!=\frac{1}{\sqrt{2}}(-|u\bar{u}\rangle+|d\bar{d}
\rangle$&\\
\hline
\end{tabular}
\end{center}
\label{structure}
\end{table}
In Table \ref{structure} the quantity $\Gamma({M\to\gamma\gamma})$ is the
experimental  two-photon decay width.  
From Table \ref{structure} it can be seen that the $q\bar{q}$ structures 
adopted in Eqs. (\ref{struc1}) to
 (\ref{struc3}) are confirmed by the two-photon widths
$\Gamma({M\to\gamma\gamma})$ to a good or at least reasonable
approximation. Furthermore, it is confirmed that the $\eta$ and $\eta'$ mesons
have  the $q\bar{q}$ structures 
\begin{equation}
|\eta\rangle \approx \frac{1}{\sqrt{2}}\left( \frac{u\bar{u}
+d\bar{d}}{\sqrt{2}}-ss \right), \quad
|\eta'\rangle \approx \frac{1}{\sqrt{2}}\left( \frac{u\bar{u}
+d\bar{d}}{\sqrt{2}}+ss \right).
\label{etstruc}
\end{equation}

\section{Approaches to  mass predictions for scalar mesons}

The prediction of the structures and the masses of scalar mesons has attracted 
many researchers.
This is especially true for the $f_0(980)$ and $a_0(980)$ mesons. In 
most of these
approaches potential models and the interplay between $(q\bar{q})^2$
states and the $K\bar{K}$ channel plays a major role 
(see e.g. \cite{weinstein90,beveren06,buck10} and references therein).

In the following we supplement on these considerations by investigating the
effects of 
spontaneous (dynamical) and explicit symmetry breaking. Here we use the term
spontaneous symmetry breaking in connection with the linear $\sigma$ model
(L$\sigma$M) 
and dynamical symmetry breaking in connection with the Nambu--Jona-Lasinio
(NJL) model. Similar approaches have been discussed at an early stage
of the development \cite{scadron82}.

\subsection{Dynamical symmetry breaking in the light-quark sector}

The mass generation of scalar mesons is well investigated in the light-quark 
sector where we have the $SU(2)$ linear $\sigma$ model, the Nambu--Jona-Lasinio
(NJL) model and the bosonized version of the NJL  model. The bosonized NJL
model is essentially equivalent to the quark-level linear $\sigma$ model
(QLL$\sigma$M) of Delbourgo and Scadron \cite{delbourgo95,beveren09}. 
Because of the importance of
these three models for the further investigations  we give a brief description 
 and the main results in the following.

For two flavors the Lagrangian of the Nambu--Jona-Lasinio (NJL) model 
has been formulated in two equivalent ways 
\cite{lurie64,eguchi76,vogl91,klimt90,klevansky92,hatsuda94}
\begin{eqnarray}
&&{\cal L}_{\rm NJL}=\bar{\psi}(i\fpartial-m_0) \psi
+ \frac{G}{2}[(\bar{\psi}\psi)^2+(\bar{\psi}i\gamma_5\vectau\psi)^2],
\label{NJL1}\\
{\rm and}\hspace{1cm} && \nonumber\\
&&{\cal L'}_ {\rm
  NJL}=\bar{\psi}i\fpartial\psi-g\bar{\psi}(\sigma+i\gamma_ 5
\vectau\cdot\pibf)\psi-\frac12\delta\mu^2(\sigma^2+\pibf^2)+\frac{gm_0}{G}
\sigma,
\label{NJL2}\\
{\rm where} \hspace{1cm}&&\nonumber\\
&&G=g^2/\delta\mu^2\quad \mbox{and}\quad \delta\mu^2=(m^{\rm cl}_\sigma)^2.
\label{grelations}
\end{eqnarray}
Eq. (\ref{NJL1}) describes the four-fermion version of the
NJL model and Eq.  (\ref{NJL2}) the bosonized  version. The quantity
$m_0=(m^0_u+m^0_d)/2$ is the average current quark mass.
The quantity $\psi$ denotes the spinor of constituent quarks with two
flavors. The quantity $G$ is the coupling constant of the four-fermion
version, $g$ the Yukawa coupling constant and $\delta\mu$ a mass parameter
entering into the mass counter-term of Eq. (\ref{NJL2}). The coupling
constants $G$, $g$ and  the mass parameter $\delta\mu$ 
are related to each other 
and to the $\sigma$ meson mass in the chiral limit (cl),
$m^{\rm cl}_\sigma$, as
given in (\ref{grelations}). 

Using diagrammatic techniques the following  equations may be found 
\cite{klevansky92,hatsuda94}  for the non-strange $(\pibf,\sigma)$ sector
\begin{eqnarray}
&&M^*=m_0+ 8\, i\, G N_c \int^{\Lambda}\frac{d^4 p}{(2\pi)^4}
\frac{M^*}{p^2-M^{*2}},\quad M=-\frac{8\,iN_cg^2}{(m^{\rm cl}_\sigma)^2}
\int\frac{d^4p}{(2\pi)^4}\frac{M}{p^2-M^2},
\label{gapdiagram}\\
&&f^2_\pi = -4\,i\,N_cM^{*2} \int^{\Lambda}\frac{d^4p}{
(2\pi)^4}\frac{1}{(p^2-M^{*2})^2},\quad f_0=-4iN_cgM\int\frac{d^4p}{(2\pi)^4}
\frac{1}{(p^2-M^2)^2},
\label{fpiexpress}\\
&&m^2_\pi=-\frac{m_0}{M^*}\frac{1}{4\,i\,G N_c I(m^2_\pi)},\quad
I(k^2)=\int^{\Lambda}\frac{d^4p}{(2\pi)^4}\frac{1}{[(p+\frac12 k)^2-M^{*2}][
(p-\frac12 k)^2-M^{*2}]}.
\label{pionmass-2}
\end{eqnarray}
The expression given on the l.h.s. of (\ref{gapdiagram}) is the 
gap equation with $M^*$
being the mass of the constituent quark with the contribution 
$m_0$ of the
current quarks included. The r.h.s.
shows the gap equation for the constituent quark mass $M$ in the chiral limit.
The l.h.s. of Eq.
(\ref{fpiexpress}) represents  the pion decay constant and the r.h.s.
the same quantity in the chiral limit. 
The expression given in Eq.   (\ref{pionmass-2}) is a generalized version
of the Gell-Mann--Oakes--Renner (GOR) relation
\begin{equation}
f^2_\pi \, m^2_\pi=-\frac12 (m^0_u+m^0_d)\langle \bar{u}u+\bar{d}d \rangle,
\label{gelloare}
\end{equation}
where $m^0_u$ and $m^0_d$ are the current-quark masses of the $u$ and $d$
quark, respectively. For further details we refer to 
\cite{schumacher06,klevansky92}.

Making use of dimensional regularization the
Delbourgo-Scadron \cite{delbourgo95} relation 
\begin{equation}
M=\frac{2\pi}{\sqrt{N_c}}f_0, \quad N_c=3
\label{sigmamass}
\end{equation}
may be obtained from 
the r.h.s of Eqs. (\ref{gapdiagram}) and (\ref{fpiexpress}).
This important relation shows that the mass of the constituent quark 
in the chiral limit and the pion decay constant in the chiral limit are
proportional to each other. This relation is valid independent of the
flavor content of the constituent quark, e.g. also for a constituent quark
where the  $d$-quark is replaced by an $s$ quark. Furthermore, it has been
shown \cite{delbourgo98,delbourgo02} that (\ref{sigmamass}) is  valid independent of the
regularization scheme.
Then 
with   the pion decay constant in the chiral limit $f_0=89.8$ MeV
\begin{equation}
m^{\rm cl}_\sigma=2\,M=652\,\,\text{MeV}
\label{chirallimit1}
\end{equation}
can be 
derived \cite{delbourgo95,schumacher06}.
  With the average pion mass $\hat{m}_\pi=138$ MeV inserted into 
\begin{equation}
m^2_{\sigma}=(2\,M)^2+\hat{m}^2_\pi,
\label{chirallimit2}
\end{equation}
the $\sigma$ meson mass is predicted  in the QLL$\sigma$M to be 
\begin{equation}
m_{\sigma}= 666\, \text{MeV}
\label{chirallimit3}
\end{equation}
as given already in Eq. (\ref{width3}).

The value given in (\ref{chirallimit3}) is the most frequently cited
``standard'' mass of the $\sigma$-meson as predicted by the QLL$\sigma$M.
This value implies that explicit symmetry breaking due to non-zero
current-quark masses enters into the $\sigma$-meson mass through the pseudo
Goldstone boson mass $\hat{m}_\pi$ only, whereas the mass $M$ of the
constituent quark is not modified by explicit symmetry breaking.

This result may be compared with  the predictions of the NJL model
\cite{klevansky92}. Making use of the l.h.s. of Eq.(\ref{fpiexpress})
and of Eq. (\ref{sigmamass}) we arrive at
\begin{equation}
\frac{f^2_\pi}{M^{*2}}=\frac{N_c}{4\pi^2}\frac{\int\frac{d^4p}{(2\pi)^4}
\frac{1}{(p^2-M^{*2})^2}}{\int\frac{d^4p}{(2\pi)^4}\frac{1}{(p^2-M^2)^2}}.
\label{eq41}
\end{equation}
Applying dimensional regularization in the form \cite{thomas00} 
\begin{equation}
\int\frac{d^Dk}{(2\pi)^D}\frac{1}{(k^2-m^2+i\,\epsilon)^2}
=i\frac{(m^2)^{-\epsilon}}{(4\pi)^{2-\epsilon}}\frac{\Gamma(\epsilon)}
{\Gamma(2)},\quad D=4-2\epsilon
\label{eq42}
\end{equation}
 instead of
regularization through a cut-off $\Lambda$
we arrive at
\begin{equation}
\frac{f^2_\pi}{M^{*2}}=\frac{N_c}{4\pi^2}\left(\frac{M^{*2}}{M^2}
\right)^{-\epsilon} \rightarrow\frac{N_c}{4\pi^2}
\label{eq43}
\end{equation}
for $\epsilon\rightarrow 0$.
This means that the expression given in Eq. (\ref{sigmamass}) 
 valid in the chiral limit
can be transferred to the case where
explicit symmetry breaking is included,
 leading to
\begin{equation}
M^*\equiv  M^*_{(u,d)} = \frac{2\pi}{\sqrt{3}}f_\pi.
\label{eq43}
\end{equation}
In the NJL model the relation  for the mass of the $\sigma$ meson in 
the presence of explicit 
symmetry breaking is given in the form \cite{klevansky92}
\begin{equation}
m^2_\sigma=4M^{*2}_{(u,d)}+\hat{m}^2_\pi.
\label{pionsigmalink}
\end{equation}
Making use of $f_\pi=92.42\pm0.26$ MeV we arrive at
\begin{equation}
m_\sigma=685\,\,\text{MeV}.
\label{sigmassbig}
\end{equation}
The $\sigma$ meson mass given in (\ref{sigmassbig}) is larger than the one
given in (\ref{chirallimit3}) by less than 3\%. Therefore, for most of the
applications this  difference is not of relevance. However, in case of
a strange-quarks content in the constituent quark the effects of explicit
symmetry breaking become sizable. This can be shown by repeating the 
arguments given above in a two-flavor theory where the $d$ quark is replaced 
by a $s$ quark. In this case we have
\begin{equation}
M^*_{(u,s)}=\frac{2\pi}{\sqrt{3}}f_K
\label{eq47}
\end{equation}
where $M^*_{(u,s)}$ is the mass of a constituent quark with an equal number of 
$u$ quarks and $s$ quarks. For a meson with an equal number of $u$ quarks 
and $s$ quarks this leads to the mass relation
\begin{equation}
m^2_{(u,s)}=4M^{*2}_{(u,s)}+m^2_K.
\label{eq48}
\end{equation}

\subsection{Spontaneous symmetry breaking in a
$SU(2)$ linear $\sigma$ model (L$\sigma$M) }

\begin{figure}[h]
\begin{center}
\includegraphics[width=0.4\linewidth]{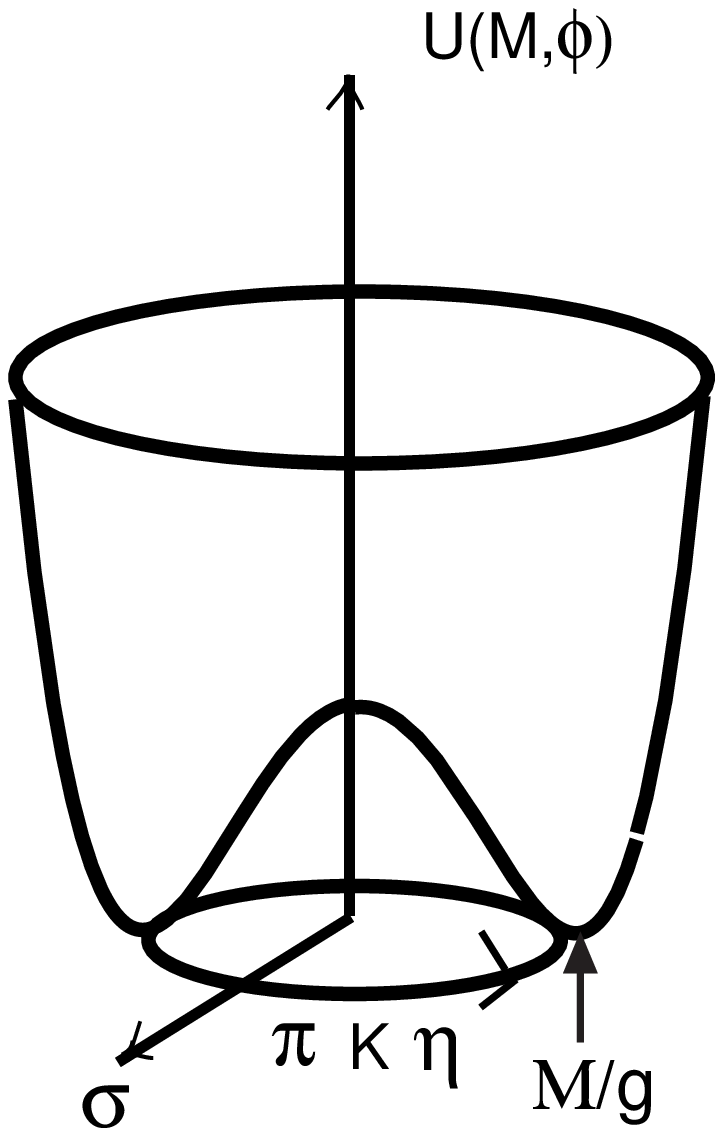}
\includegraphics[width=0.5\linewidth]{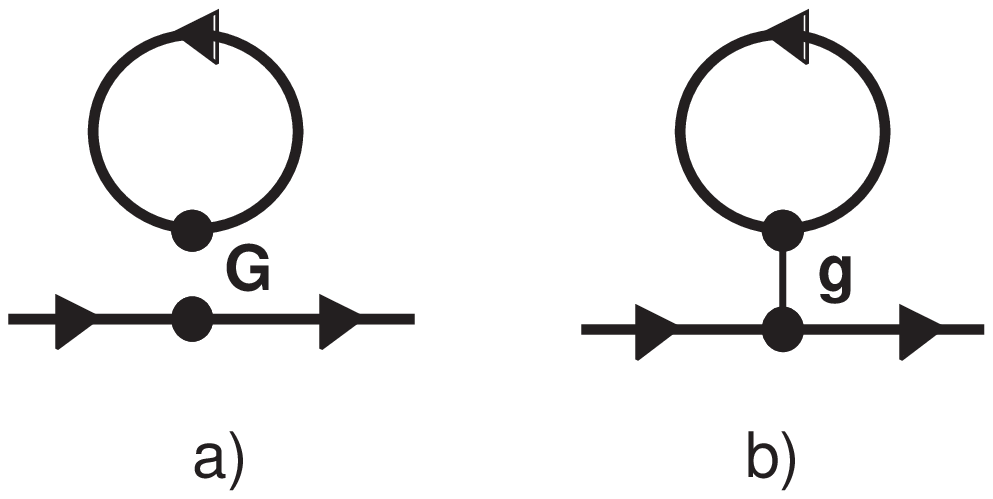}
\end{center}
\caption{Left panel: Spontaneous symmetry breaking in the chiral limit
illustrated by the  L$\sigma$M: In the $SU(2)$ sector there is one 
``strong Higgs boson'', 
the  $\sigma$ meson having a mass of $m^{\rm cl}_\sigma=652$ MeV taking 
part in spontaneous  symmetry breaking, accompanied by an isotriplet
of massless $\pi$ mesons serving as Goldstone bosons. In the $SU(3)$ sector
there are  8 massless Goldstone bosons $\pibf$, $K$, $\eta$, and nine 
scalar mesons
$\sigma$, $\kappabf$, $f_0$ and $a_0$, all of them 
having the same mass as the $\sigma$ meson in the chiral limit. 
The mass degeneracy is removed
by explicit symmetry breaking.
Right panel: Tadpole graphs of chiral
symmetry breaking. a) Four fermion version of the Nambu-Jona--Lasinio (NJL)
model, b) bosonized NJL model.}
\label{mexicanhat}
\end{figure}
\begin{figure}[h]
\begin{center}
\includegraphics[width=0.5\linewidth]{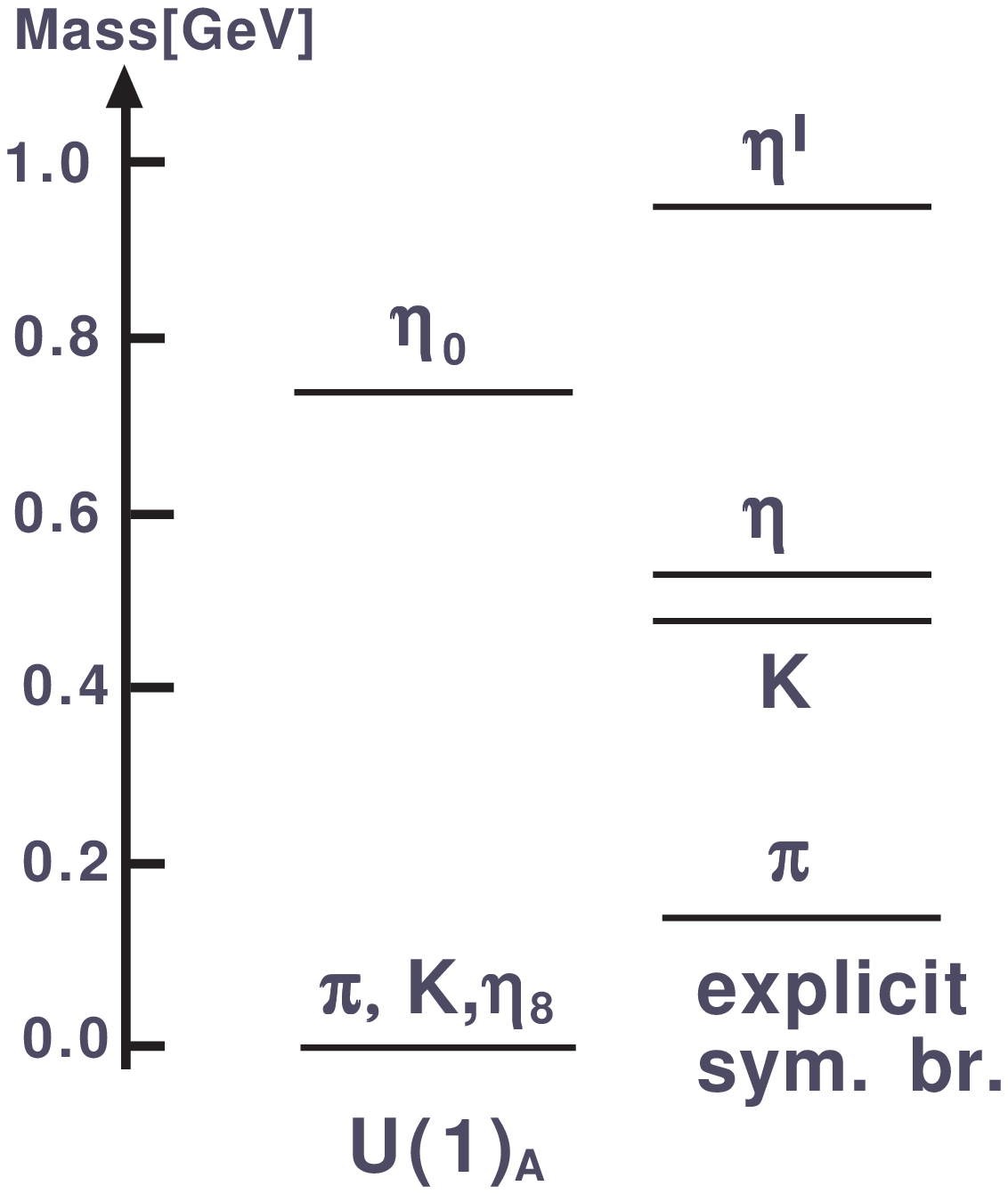}
\end{center}
\caption{Pseudoscalar mesons after U(1)$_A$ symmetry breaking (left column)
and after additional explicit symmetry breaking (right column).}
\label{levelscheme}
\end{figure}
The L$\sigma$M \cite{schwinger57,gellmann60,dealfaro73} is 
the first  theory of  symmetry breaking
in particle physics,
explaining the mass of  the $\sigma$ meson and of the constituent quarks. 
The underlying mechanism is spontaneous symmetry breaking due to which
the $\sigma$ field obtains a vacuum expectation value of $\langle 0|\sigma
|0\rangle =f_0$ in the chiral limit.
Later on spontaneous symmetry breaking has  been adopted to the electroweak 
sector  where it describes
 the mass generation of the Higgs boson and the related mass generations
of current quarks, leptons and electroweak gauge bosons.

For our approach it is of advantage to first outline the common  case 
\cite{donoghue92,dealfaro73} of the $SU(2)$ linear $\sigma$ model and to
supplement information obtained from the NJL model. 
In the chiral limit the Lagrangian may be written in the form
\begin{equation}
{\cal L}=\frac12\partial_\mu\pibf\cdot\partial^\mu
\pibf + \frac12 \partial_\mu\sigma\partial^\mu\sigma 
+\frac{\mu^2}{2}(
\sigma^2+\pibf^2)-\frac{\lambda}{4}(\sigma^2+\pibf^2)^2.
\label{lagrang1}
\end{equation}
For $\mu^2>0$ and $\lambda>0$, the model exhibits the phenomenon of
spontaneous symmetry breaking which will be explained in the following
 in more detail.

 We 
infer from the sigma model of Eq. (\ref{lagrang1}) the potential energy 
\begin{equation}
V(\sigma,\pibf)=-\frac{\mu^2}{2}(\sigma^2+\pibf^2)+\frac{\lambda}{4}
(\sigma^2+\pibf^2)^2.  
\label{lagrang6}
\end{equation}
Minimization of $V(\sigma,\pibf)$ 
reveals the set of degenerate ground states to be those with
\begin{equation}
\sigma^2+\pibf^2=\frac{\mu^2}{\lambda}.
\label{lagrang7}
\end{equation}
Of these we select the particular ground state 
\begin{equation}
\langle\sigma\rangle_0=\sqrt{\frac{\mu^2}{\lambda}}\equiv v,
\quad \langle \pibf \rangle_0=0,
\label{lagrang8}
\end{equation}
where $v$ is the vacuum expectation value (VEV) of the $\sigma$ field
in the chiral limit which can be shown to be given by the pion decay
constant $f_0$ in the chiral limit:
\begin{equation}
v=f_0.
\label{lagrang9}
\end{equation}
The mexican hat potential is shown in Figure \ref{mexicanhat} together
with the graphs describing symmetry breaking in the four-fermion NJL model
and the bosonized  NJL model. In Figure \ref{mexicanhat} use has been made
of the Goldberger-Treiman (GT) relation on the quark level and in the chiral
limit 
\begin{equation}
M=gf_0
\label{lagrang10}
\end{equation}
where $M=326$ MeV  is the constituent quark mass in the chiral 
limit and where 
the coupling constant $g$ on the quark level is given by the 
Delbourgo-Scadron relation
\begin{equation}
g=\frac{2\pi}{\sqrt{N_c}},\quad N_c=3.
\label{lagrang11}
\end{equation}
Then the further evaluation of the L$\sigma$M and NJL models in 
the chiral limit
leads to \cite{dealfaro73}
\begin{equation}
\mu=\sqrt{2}M=461\,\,\text{MeV}\,\,\,\text{and} \,\,\, 
\lambda=2g^2=\frac{8\pi^2}{3}=26.3.
\label{lagrang12}
\end{equation}

The $SU(2)_l\times SU(2)_R$ symmetry of the sigma model is explicitly
broken if the potential $V(\sigma,\pibf)$ is made slightly asymmetric, 
{\it e.g.} by the addition of the term
\begin{equation}
{\cal L}_{\rm breaking}=a\sigma
\label{lagrang13}  
\end{equation}
to the basic Lagrangian of Eq. (\ref{lagrang1}). To first order in the
quantity $a$, this shifts the minimum of the potential to
\begin{equation}
v=\sqrt{\frac{\mu^2}{\lambda}}+\frac{a}{2\mu^2}
\label{lagrang14}
\end{equation}
where 
\begin{equation}
a=f_\pi m^2_\pi.
\label{lagrang15}
\end{equation}
This leads to \cite{dealfaro73}
\begin{equation}
m^2_\sigma=2\lambda f^2_\pi +m^2_\pi=\frac{16\pi^2}{3}f^2_\pi+\hat{m}^2_\pi,
\label{lagrang16}
\end{equation}
and  to $m_\sigma=685$ MeV as derived before.

\subsection{The $SU(3)$ NJL model}

An essential difference between the $SU(2)$ and $SU(3)$ sectors  is that 
symmetry
breaking due to the $U(1)_A$ anomaly has to be taken into account in the
latter. As shown in Figure \ref{levelscheme} this effect is quite sizable in
case of the $\eta_0-\eta_8$ mass splitting.

\begin{figure}[h]
\begin{center}
\includegraphics[width=0.5\linewidth]{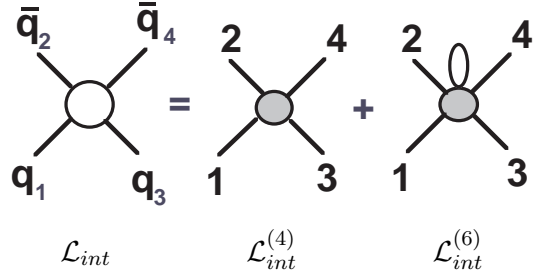}\\
${\cal L}_{int}\hspace{18mm}{\cal L}^{(4)}_{int}\hspace{18mm}{\cal L}^{(6)}_{int}$
\end{center}
\caption{Graphs representing the interaction Lagrangian of the $SU(3)$ NJL
  model.
The six-point interaction term ${\cal L}^{(6)}_{int}$ is approximated by an
  effective four-point interaction term. 
 } 
\label{nambu}
\end{figure}

The Lagrangian is
\begin{equation}
{\cal L}_{NJL}=\bar{\Psi}(i\fpartial-m_0)\Psi+{\cal L}_{int.}
\label{su3Lag1}
\end{equation}
with the current quark mass matrix $m_0={\rm diag}(m^0_u,m^0_d,m^0_s)$
( see \cite{vogl91,klimt90,klevansky92,hatsuda94} and references therein).
The interaction part
\begin{equation}
{\cal L}_{int}={\cal L}^{(4)}_{int}+{\cal L}^{(6)}_{int}
\label{su3Lag2}
\end{equation}
has a local four-point interaction ${\cal L}^{(4)}_{int}$ and a
$U(1)_A$-breaking term ${\cal L}^{(6)}_{int}$ which is minimally a six-point
interaction (see Figure \ref{nambu}). 
The four-point interaction  term may be written in the form
\begin{equation}
{\cal L}^{(4)}_{int}=\frac{G_S}{2}\sum^8_{i=0}[(\bar{\psi}\lambda^i\psi)^2
+(\bar{\psi}i\gamma_5 \lambda^i \psi)^2]
\label{lg1}
\end{equation}
and the six-point interaction term in the form
\begin{equation}
{\cal L}^{(6)}=\frac{G_D}{2}\{{\rm det}[\bar{\psi}(1+\gamma_5)\psi]
+{\rm det}[\bar{\psi}(1-\gamma_5)\psi]\}.
\label{lg2}
\end{equation}
It can be written in the form \cite{vogl91}
\begin{equation}
{\cal L}^{(6)}=\frac{G_D}{12}d_{ijk}\left[ \frac13 
(\bar{\psi}\lambda^i\psi)(\bar{\psi}\lambda^j\psi)
+(\bar{\psi}\gamma_5\lambda^i\psi)(\bar{\psi}\gamma_5\lambda^j\psi)
\right](\bar{\psi}\lambda^k\psi).
\label{lg3}
\end{equation}
The $d_{ijk}$ are the symmetric $SU(3)$ structure constants;
$d_{000}=\sqrt{\frac23}$  and $d{0jk}=-\sqrt{\frac16}$ for $j,k,\neq 0$.       
This six-point interaction term  can be decomposed \cite{klevansky92}
into terms that are
proportional 
\begin{equation}
(\bar{\psi}\lambda^i \psi)^2 \quad \text{and} \quad 
(\bar{\psi}\gamma_5 \lambda^i \psi)^2.
\label{lg3}
\end{equation}
These terms are of the same structure as the terms provided by the 
four-point interaction term. But in addition one has {\cal mixed}
terms as 
\begin{equation}
(\bar{\psi}\lambda^0\gamma_5\psi)(\bar{\psi}\lambda^8\gamma_5\psi),\,\,
(\bar{\psi}\lambda^8\gamma_5\psi)(\bar{\psi}\lambda^0\gamma_5\psi)
\label{lg4}
\end{equation}
and their scalar counterparts
\begin{equation}
(\bar{\psi}\lambda^0\psi)(\bar{\psi}\lambda^8\psi),\,\,
(\bar{\psi}\lambda^8\psi)(\bar{\psi}\lambda^0\psi).
\label{lg5}
\end{equation}
The terms given in (\ref{lg3}) may be treated as part of the four-point
interaction. This leads to the conclusion that not $G_S$ alone but a linear
combination of $G_S$ and $G_D$ corresponds to an effective four-point
interaction. This linear combinations is derived in the following paragraph. 
The pseudoscalar {\cal mixed} terms in Eq. (\ref{lg4}) lead to the well-known
mass splitting of the pseudoscalar mesons $\eta_0$ and $\eta_8$ present in the
chiral limit which remains to be essentially unmodified for
the physical mesons $\eta$ and $\eta'$ but - at the present
status of the discussion - it 
remains unknown whether or not there is also a mass splitting
due to the six-point interaction in case of scalar mesons. It is the purpose
of the following discussion to clarify this important  question.

In practical applications to the pseudoscalar and scalar meson sectors the 
model makes use of the two coupling parameters $G_S$ and $G_D$ corresponding 
to ${\cal L}^{(4)}_{int}$, and ${\cal L}^{(6)}_{int}$, respectively,
such that the
pseudoscalar couplings are
 \cite{hatsuda94}
\begin{eqnarray}
&&G_\pi=G_S+G_D \langle\bar{s}s\rangle,\,\,G_{K^\pm}=G_S+G_D\langle 
d\bar{d}\rangle,\,\,G_{K^0}=G_S+G_D\langle u\bar{u}\rangle \nonumber\\
&&G_{\eta_0}=G_S-\frac23(\langle u\bar{u}\rangle +\langle d\bar{d}\rangle
+\langle s\bar{s}\rangle) G_D,\,\,
G_{\eta_8}=G_S-\frac13(\langle s\bar{s}\rangle -2\langle u\bar{u}\rangle
-2\langle d\bar{d}\rangle) G_D.
\label{G1}
\end{eqnarray} 
We now go to the chiral limit where 
$\langle u\bar{u}\rangle= \langle d\bar{d}\rangle= 
\langle s\bar{s}\rangle  =\langle q\bar{q}\rangle $. Furthermore, we
rearrange  the coupling parameters such that 
$G=G_S+G_D \langle\bar{q}q\rangle$ and  arrive at 
(see \cite{vogl91} for further justification of this step )
\begin{eqnarray}
&&G_\pi=G ,\quad  G_{K^\pm}=G,\quad   G_{K^0}=G, \nonumber\\
&&G_{\eta_0}=G-3\,G_D \langle q\bar{q}\rangle ,\quad   G_{\eta_8}=G.
\label{G2}
\end{eqnarray} 
This means that $G_{\eta_0}$ differs from the coupling parameters $G$ 
of the other pseudoscalar mesons by  the positive 
amount of $-3G_D\langle q\bar{q}\rangle$. In the chiral limit all the 
pseudoscalar mesons have zero mass as appropriate  for Goldstone bosons
with the exception of $\eta_0$  which gets a mass through $U(1)_A$ 
symmetry breaking which is given here by the six quark interaction term.
Apparently this effect is quite sizable as can be seen in Figure 
\ref{levelscheme}.

Now we come to  scalar mesons and as a  first step describe the flavor
wave functions by the same $SU(3)$ expressions as in case of the pseudoscalar
mesons. This means that the pseudoscalar wave-functions
\begin{equation}
\eta_0=\frac{1}{\sqrt{3}}(u\bar{u}+d\bar{d}+s\bar{s})^1S_0,\quad
\eta_8=\frac{1}{\sqrt{6}}(u\bar{u}+d\bar{d}-2s\bar{s})^1S_0
\label{pseu}
\end{equation}
have the scalar analogs
\begin{equation}
\epsilon_0=\frac{1}{\sqrt{3}}(u\bar{u}+d\bar{d}+s\bar{s})^3P_0,\quad
\epsilon_8=\frac{1}{\sqrt{6}}(u\bar{u}+d\bar{d}-2s\bar{s})^3P_0.
\label{sca}
\end{equation}
The scalar analogs of the coupling parameters 
in Eq. (\ref{G1}) only differ in the sign of $G_D$
\cite{vogl91,klimt90,klevansky92,hatsuda94}. In the chiral limit this leads to 
\begin{eqnarray}
&&G_{\delta}=G ,\quad G_{\kappa^\pm}=G,\quad G_{\kappa^0}=G, \nonumber\\
&&G_{\epsilon_0}=G+3\,G_D \langle q\bar{q}\rangle ,\quad G_{\epsilon_8}=G,
\label{G2}
\end{eqnarray} 
where use has been made of $G=G_S-G_D\langle q\bar{q}\rangle$.
This means that we obtain a mass splitting between $\epsilon_0$,
the scalar analogs of $\eta_0$,  and
$\epsilon_8$, the scalar analog of $\eta_8$, such that  $\epsilon_0$
differs by a negative 
mass term from  all the other scalar mesons. Following the arguments
outlined in connection with the pseudoscalar mesons this would mean
that in the chiral limit all the scalar mesons $\kappa$, $f_0$ and $a_0$
would have the mass of $2M=652$ MeV and $\epsilon_0$
a smaller mass due to the  effects of the $U(1)_A$ symmetry breaking. 
Up to this point  the $\epsilon_0$ meson has been used to represent
the  $\sigma$ meson in the chiral limit
which, according to subsection 4.1, does not show any 
dependence on $U(1)_A$ symmetry breaking. Apparently, here we find a 
difference from
the pseudoscalar case where the effects of $U(1)_A$ symmetry breaking 
predicted for the
$\eta_0$ meson is found in the physical $\eta'$ meson, though the stuctures
of the two mesons are close to each other but not identical. This difference
between the pseudoscalar and the scalar case 
comes not as a surprise because even in the chiral limit the $\epsilon_0$ 
and the $\epsilon_8$
do not represent the flavor structures  of the $\sigma$ and the $f_0(980)$
mesons. The $^3P_0$ components accompanying the $(q\bar{q})^2$ structures are
\begin{equation}
\epsilon'_0=\frac{u\bar{u}+d\bar{d}}{\sqrt{2}},\quad
\epsilon'_8=\frac{1}{\sqrt{2}}\left(\frac{u\bar{u}+d\bar{d}}{\sqrt{2}}
-s\bar{s}\right)
\label{epsprime}
\end{equation}
which do not have the necessary symmetry among the three flavors in order to
make $U(1)_A$ symmetry breaking effective.
Summarizing we can state that the only effect
of $SU(1)_A$ symmetry breaking is to project the $\eta'$ meson out of the
number of pseudo Goldstone bosons.

\subsection{The $SU(3)$ L$\sigma$M}

The transition
from two flavors to three flavors is described in many recent papers
\cite{schechter71,tornqvist99,tornqvist02,wu04,parganlija10,chen10}.
The scalar nonet is put into the hermitian part of a $3\times 3$ matrix 
$\Phi$ and a pseudoscalar nonet into the anti-hermitian part of $\Phi$.
One has (for the notation used here see  \cite{tornqvist02})
\begin{equation}
\Phi=S+iP=\sum^8_{a=0}(\sigma_a+ip_a)\lambda_a/\sqrt{2},
\label{three1}
\end{equation}
where $\lambda_a$ are the Gell-Mann matrices, and 
$\lambda_0=\sqrt{\frac23}1\!\!1$. Then the potential is
\begin{equation}
V(\Phi)=-\frac12\mu^2{\rm Tr}[\Phi\Phi^\dagger]
+\lambda{\rm Tr}[\Phi\Phi^\dagger\Phi\Phi^\dagger]
+\lambda'({\rm Tr}[\Phi\Phi^\dagger])^2+{\cal L}_{SB},
\label{three2}
\end{equation}
where $\lambda'$ is a small parameter compared to $\lambda$ and 
${\cal L}_{SB}$ contains an explicit symmetry breaking term and an
$U_A(1)$ breaking term $\propto({\rm det}\Phi+{\rm det}\Phi^\dagger)$.
There are different proposals for a further evaluation of this ansatz 
and in this connection we refer to the following works 
\cite{tornqvist99,tornqvist02,wu04,parganlija10}. 
The problem encountered in these approaches is to get a physically justified
criterion for the treatment of the $U(1)_A$ breaking term. 
Following \cite{tornqvist99} we write down the symmetry breaking terms in the
most simple form
\begin{equation}
{\cal L}_{SB}=\epsilon_\sigma\, \sigma_{u\bar{u}+d\bar{d}}
+\epsilon_{s\bar{s}}\,\sigma_{s\bar{s}}+\beta({\rm det}\Phi
+{\rm det}\Phi^\dagger)
\label{symbreak}
\end{equation}
where $\epsilon_\sigma=m^2_\pi f_\pi,\,\,\epsilon_{s\bar{s}}
=(2m^2_Kf_K-m^2_\pi 
f_\pi)/\sqrt{2}$. 

We have given the relations (\ref{three1}) - (\ref{symbreak}) here 
as one example
of the different approaches  proposed in  
\cite{schechter71,tornqvist99,tornqvist02,wu04,parganlija10,chen10}.
The common goal of  these approaches is to obtain relations for the masses
of the
scalar and pseudoscalar mesons in terms of  parameters, especially
the parameter $\beta$ in   (\ref{symbreak})    which describes the strength
of the $U(1)_A$ symmetry breaking effect. In the framework of
the $SU(3)$ L$\sigma$M alone the factor $\beta$
 is predicted to
have an influence on the masses of most of the scalar and pseudoscalar mesons.
It is the strategy of the present paper not to  exclusively rely on one
theoretical ansatz but to take into account in a phenomenological way all the
available information. Differences between the present and previous approaches
and results are due to this  difference in the strategy and are not in conflict
otherwise. 

The elaborate investigation in the  subsection 4.3 has made quite
clear that  the  only effect of  $U(1)_A$ symmetry breaking is to project
the $\eta'$ meson out of the number of pseudo Goldstone bosons, thus making
it unnecessary to  take into account this effect after the $\eta'$
mesons has been removed from the list of particles to be considered. 
Furthermore, there is no reason to take into account the small flavour mixing
effect due to the term proportional $\lambda'$ in (\ref{three2}).
But there is agreement between the present approach and the approach
described in (\ref{three1}) - (\ref{symbreak})  that explicit symmetry 
breaking can be expressed
through terms of the form
$m^2_\pi f_\pi$ and $m^2_Kf_K$ or appropriate linear combinations of
$m^2_\pi f_\pi$ and $m^2_Kf_K$. This is the basis of the phenomenological
approach described on the next subsection.

\subsection{Prediction of masses of the scalar mesons}

The $SU(3)$ L$\sigma$M as treated in the foregoing subsection
has the disadvantage that explicit use is made of a $q\bar{q}$ structure
of scalar mesons. Since we know that scalar mesons are mainly of a
tetra-quark structure with only a minor $q\bar{q}$ contribution we have to
look for a structure-independent treatment of chiral symmetry breaking.
For the $SU(2)$ L$\sigma$M this idea is not new 
(see e.g. \cite{abdelrehim03})
because the Lagrangian written
down in  (\ref{lagrang1}) is formulated in terms of fields rather than in
terms of particles with a definite $q\bar{q}$ structure. Another approach
to the prediction of masses of scalar mesons  based on dynamical symmetry
breaking has been  described in subsection 4.1. As a result of those 
considerations we may state that in the chiral limit all the 
scalar mesons considered in this
work  have a mass which is equal to $m^{\rm cl}_\sigma =652$ MeV independent
of the special flavour structure. Differences in the masses occur only due to
the effects of explicit symmetry breaking. The rules according to which 
explicit symmetry breaking modifies the masses has partly already been
investigated in subsection 4.1, so that in the present subsection only some 
amendments are necessary.

In the present paragraph we present a discussion which is based on
spontaneous symmetry breaking and supplement the results by arguments based on
dynamical symmetry breaking where necessary. 
It is  possible to write down the  Lagrangian given in (\ref{lagrang1}) 
in the general form
\begin{equation}
{\cal L}_\sigma=\frac12[(\partial_\mu \phi_1)^2+(\partial_u \phi_2)^2
+\mu^2(\phi^2_1+\phi^2_2)]-\frac14\lambda(\phi^2_1+\phi^2_2)^2+a_\sigma \phi_1.
\label{lagr1}
\end{equation}
where $\phi_1$ corresponds to the scalar and $\phi_2$ to pseudoscalar
component. 
The last term in (\ref{lagr1}) takes into account explicit symmetry breaking
where $a_\sigma=m^2_\pi f_\pi $. 
For the generalization of Eq. (\ref{lagr1})  it is necessary to find 
Goldstone-boson (GB)
partners of the scalar mesons $\kappa(800)$, $f_0(980)$ and $a_0(980)$
taking into account the $(q\bar{q})^2$ structure of the scalar
mesons. For this purpose we define the quantity f$_s$ as the fraction of
strange quarks and/or antistrange quarks in the four-quark configuration of
the scalar meson.
Using the wave functions from Table \ref{qqqq}, we may easily evaluate
the value of f$_s$. 
\begin{table}[h]
\caption{Fraction $f_s$ of strange quarks and/or antistrange quarks in the  
$(q\bar{q})^2$ 
structure of scalar mesons and
  Goldstone bosons (GB) with the same fraction of strange quarks}
\begin{center}
\begin{tabular}{lcl}
\hline
$(q\bar{q})^2$ meson& f$_s$ & GB\\
\hline
$\sigma$& 0 & $\pi$\\
$\kappa(800)$ & $\frac14$ & $K,\pi$\\
$f_0(980)$, $a_0(980)$ & $\frac12$ & $\eta$ \\
\hline
\end{tabular} 
\label{tabyB}
\end{center}
\end{table}
Making the reasonable assumption that the scalar meson and the GB partner 
should have the same strange-quark fraction f$_s$ we arrive at the $\pi$ 
meson as the GB partner of the $\sigma$ meson what is well known and a 
combination of $K$ and $\pi$ for the $\kappa(800)$ and the $\eta$ for
$a_0(980)$ and $f_0(980)$. These results are given in Table \ref{tabyB}.
In the framework of the Lagrangian given in Eq. 
(\ref{lagr1})  scalar and pseudoscalar mesons 
may be represented by a complex field
\begin{equation}
\Phi_\sigma=\frac{1}{\sqrt{2}}(\phi_1+i\,\phi_2).
\label{higgsansatz}
\end{equation}
The absence of flavor mixing makes it posible to write down  
analogous fields for the $\kappa(800)$ and $(a_0(980,f_0(980))$ sectors
leading to Lagrangians analogous 
to the one  of the L$\sigma$M as given (\ref{lagr1}). These are given 
in the form
\begin{eqnarray}
&&{\cal L}_\kappa=\frac12[(\partial_\mu \phi_3)^2+(\partial_u \phi_4)^2
+\mu^2(\phi^2_3+\phi^2_4)]-\frac14\lambda(\phi^2_3+\phi^2_4)^2+a_\kappa\phi_3,
\label{lagr2}\\
&&{\cal L}_{(a_0,f_0)}=\frac12[(\partial_\mu \phi_5)^2+(\partial_u \phi_6)^2
+\mu^2(\phi^2_5+\phi^2_6)]-\frac14\lambda(\phi^2_5+\phi^2_6)^2
+a_{(a_0,f_0)}\phi_5.
\label{lagr3}
\end{eqnarray}
The last terms in (\ref{lagr2}) and (\ref{lagr3}) take into account 
explicit symmetry breaking and are given by 
\begin{eqnarray}
&&a_\kappa=\frac12 (m^2_K  f_K + m^2_\pi   f_\pi ), \label{three4}\\
&&a_{(a_0,f_0)}=  m^2_\eta  f_\eta. \label{three5}
\end{eqnarray}
Since the $K$  and $\eta$ mesons have about equal masses we expect
$f_\eta\approx f_K$. This conclusion follows from considerations contained in 
\cite{sanzcillero04} where the relation between mass and  decay constant of 
pseudoscalar mesons is investigated.
\begin{figure}[h]
\begin{center}
\includegraphics[width=0.5\linewidth]{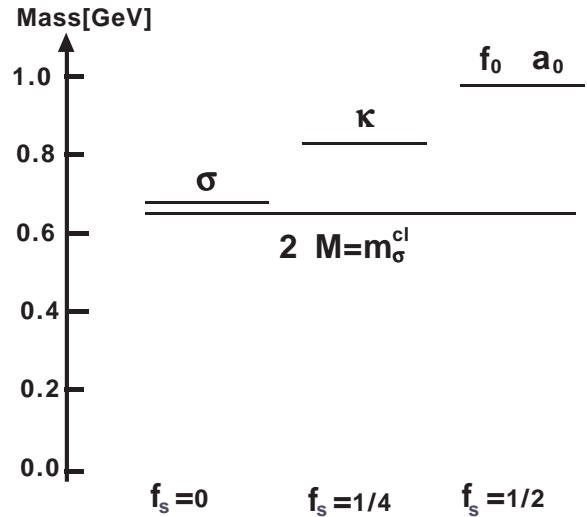}
\end{center}
\caption{Predicted masses of scalar mesons. Long horizontal line: Mass of
  $\sigma$ meson $m^{\rm cl}_\sigma$ in the chiral limit (cl). M is the
constituent quark mass in the chiral limit. $f_s$ is the fraction of strange
quarks and/or antistrange quarks in the four-quark configuration of the scalar
meson. Effects of $U(1)_A$ symmetry breaking are negligible in comparison with 
$2M=m^{\rm cl}_\sigma$. 
 } 
\label{smass}
\end{figure}
The further evaluation proceeds in complete analogy to the case 
of the $SU(2)$ L$\sigma$M described in subsection 4.2, but now applied to the 
three cases given in (\ref{lagr1}),  (\ref{lagr2}) and  (\ref{lagr3}).
This consideratiom leads to the three mass formulae
\begin{eqnarray}
&&m^2_\sigma=\frac{16\pi^2}{3}f^2_\pi+m^2_\pi,\label{three6}\\
&&m^2_\kappa=\frac{16\pi^2}{3}\frac12(f^2_\pi+f^2_K)+\frac12(m^2_\pi+m^2_K),
\label{three7}\\
&&m^2_{a_0,f_0}=\frac{16\pi^2}{3}f^2_K+m^2_\eta,
\label{three8}
\end{eqnarray}
with $f_K=113.0\pm 1.0$ MeV \cite{PDG}.
The masses  predicted in this way are $m_\sigma=685$ MeV,
$m_\kappa= 834$ MeV and $m_{a_0,f_0}= 986$ MeV in close
agreement with the experimental data. It should be noted that the mass
formulae given in (\ref{three6}) and (\ref{three8}) have already been derived
in subsection 4.1 using arguments from dynamical symmetry breaking, so that 
only the formula in (\ref{three7}) is new. The use of a linear combination
of explicit symmetry breaking terms from  the $\pi$-mesons and the $K$-meson
has already been discussed subsection 4.4 where a related procedure proposed
by T\"ornqvist \cite{tornqvist99} is discussed.

Figure \ref{smass} illustrates the result obtained. 
In the chiral limit all the scalar mesons  have the same mass given
by $m^{\rm cl}_\sigma=2\,M$, i.e. the $\sigma$ mass in the chiral limit
or, equivalently, twice the  mass of the constituent quark in the chiral limit.
Additional contributions to the mass arise due to the effects of
explicit symmetry breaking which enters into mass formulae
in two ways.  The first way is due to the fact that the pion decay constant
$f_0$ valid in the chiral limit has to be replaced by 
$f_\pi$, $f_K$ and $f_\eta$, respectively. The second way is caused by
the masses of the pseudoscalar mesons which are equal to zero in the chiral
limit. 
The predictions obtained are in a remarkable 
agreement with the experimental values.

\section{Summary and discussion}

Summarizing, we may state that the masses of scalar mesons may be calculated 
in terms of two mass components which have to be added in quadrature. The first
mass component 
$m_1=2\,M^*=\frac{4\pi}{\sqrt{3}}f_{\pi,K}$ is obtained via dynamical symmetry breaking in the 
presence of a  correction due to
explicit symmetry breaking contained in the decay constants $f_{\pi,K}$. 
The second mass component $m_2$
is due to explicit   
symmetry breaking only and may be identified with  
the mass of a pseudo-Goldstone
boson with the same fraction $f_s$ of strange quarks in the flavor 
wave-function.
There is a formal analogy with  the mass generation
in the electroweak sector, however with important differences. In the 
chiral limit we have $m^{\rm cl}_1=m^{\rm cl}_\sigma$ and 
$m^{\rm cl}_2\equiv 0$. In a 
formal way this corresponds to symmetry breaking 
in the electroweak sector with the 
$\sigma$ meson being the analog of the Higgs boson. However, there is 
no analog to the Higgs mechanism which transfers the massless Goldstone 
boson into the longitudinal component of a massive gauge field. Instead,
explicit symmetry breaking gives the Goldstone boson a mass which
becomes a part of the mass of the scalar meson. Analogous relations are found
for the scalar mesons $\kappa(800)$, $f_0(980)$ and $a_0(980)$.

In the following we write down the GOR relations as given  in \cite{thomas00}
for $\pi$ and the $K^+$ mesons and supplement it by a corresponding relation
for the $\eta$ meson.
\begin{eqnarray}
&&m^2_\pi     f^2_\pi       =-\frac12(m^0_u+m^0_d)
\langle\bar{u}u+\bar{d}d\rangle
+{\cal O}\left(({m^0_{u,d}})^2\right),\label{gell1}\\
&& m^2_{K^+}     f^2_{K^+}     =-\frac12 (m^0_u+m^0_s)
\langle\bar{u}u+\bar{s}s\rangle
+{\cal O}\left(({m^0_{s}})^2\right) \label{gell2}\\
&& m^2_{\eta}     f^2_{\eta}     =-\frac14 (m^0_u+ m^0_d +2m^0_s)
\langle\bar{u}u+\bar{s}s\rangle
+{\cal O}\left(({m^0_{s}})^2\right) \label{gell2}
\end{eqnarray}
where 
\begin{equation}
\langle\bar{u}u\rangle\simeq \langle\bar{d}d\rangle\simeq \langle\bar{s}s
\rangle\simeq -(225\,\, {\rm MeV})^3\simeq -1.5\,{\rm fm}^{-3}.
\label{upvacuum}
\end{equation}
These equations show that explicit symmetry breaking has two origins.
The first is the QCD vacuum as in case of dynamical or spontaneous symmetry
breaking. The second is the mass of the current quarks, which 
is a consequence of explicit symmetry breaking.
At this point it is possible to relate the strong and the electroweak
Higgs theories to each other by writing down
\begin{equation}
m^0_q=\frac{1}{\sqrt{2}}g_{hqq}\, v_h
\label{wein}
\end{equation}
with $v_h=(\sqrt{2}G_F)^{-1/2}\simeq 246$ GeV, as following from the
Weinberg-Salam model (see e.g. \cite{halzen84} p. 331). The quantity $g_{hqq}$
is the Higgs-quark coupling constant and $v_h$ the vacuum expectation value of
the Higgs field.

\end{document}